# Eco-driving Intelligent Systems and Algorithms

## A Patent Review

Ma, Zhipeng; Jørgensen, Bo Nørregaard; Ma, Zheng Grace

# Eco-driving Intelligent Systems and Algorithms: A Patent Review

Zhipeng Ma
*SDU Center for Energy Informatics, the Maersk Mc-Kinney Moller Institute*
*University of Southern Denmark*
Odense, Denmark
zhma@mmmi.sdu.dk

Bo Nørregaard Jørgensen
*SDU Center for Energy Informatics, the Maersk Mc-Kinney Moller Institute*
*University of Southern Denmark*
Odense, Denmark
bnj@mmmi.sdu.dk

Zheng Grace Ma
*SDU Center for Energy Informatics, the Maersk Mc-Kinney Moller Institute*
*University of Southern Denmark*
Odense, Denmark
zma@mmmi.sdu.dk

*Abstract*—The transportation industry remains a significant contributor to greenhouse gas emissions, highlighting the requirement for intelligent systems to enhance vehicle energy efficiency. The intellectual property rights of developed systems should be protected by patents. However, there is no patent overview of eco-driving intelligent systems. Unlike a scientific article, a patent documentation indicates both novelty and commercialization potential of an inventor. To address this research gap, this paper provides a patent overview of eco-driving intelligent systems and algorithms. 424 patents in the Google Patent database are analyzed. The patent analysis results show that the top three Cooperative Patent Classifications are: Y02T - climate change mitigation technologies related to transportation (50.7%), B60W - Conjoint control of vehicle sub-units of different types or different functions (34.4%) and B60L - Propulsion of electrically-propelled vehicles (20.2%). 219 patents were filed after 2016 when deep learning became popular and can be categorized into five groups: vehicle energy management, smart driving, ecological and sustainable driving, fuel consumption reduction, and driving behavior optimization. Furthermore, all 219 patents involve the physical components of the intelligent system and/or novel machine learning/deep learning algorithms. Moreover, over 70% of them are granted by the China National Intellectual Property Administration.

## I. INTRODUCTION

Anthropogenic greenhouse gases (GHG) give rise to global warming through the greenhouse effect in recent years, leading to many ecological and social problems. The transportation sector accounts for almost 30% of GHG emissions [1], about 65% of which are caused by road transport [2]. To achieve the United Nations' sustainability goals [3], the eco-driving strategies and the corresponding intelligent systems need to be studied to mitigate fuel consumption and CO2 emissions.

Eco-driving strategies involve drivers adopting optimized driving techniques, such as adjusting speed and acceleration based on road and weather conditions [4]. Additionally, planning efficient routes can minimize energy consumption by avoiding congestion and traffic signals [5].

Artificial intelligence (AI) technologies have led to the development of intelligent systems that analyze driving behavior, vehicle data, and environmental data to recommend fuel-saving practices. For instance, A CANBus (Controller Area Network)-based system is implemented in [6] to evaluate the bus drivers' eco-driving performance, resulting in a 5% reduction in energy consumption. An eco-driving assistant system is developed by H. Ma et al. [7] for guiding bus drivers to reduce fuel consumption with the decision tree algorithm. The driving data from various cars were collected by a universal OBD-II (on-board diagnostics) module and an Elman neural network framework was developed for data processing in [8], which archives eco-driving in the bus driver training.

Studies have revealed that many intelligent systems have been developed to assist in eco-driving strategies. Driving data are collected from CANBus [6] and OBD [8], and smartphones [9] are used to generate environmental data. State-of-the-art AI models are developed and applied to analyze the collected data. For example, the random forest [10] and neural networks [11] are applied to estimate fuel consumption. Furthermore, multi-agent simulation models based on the reinforcement learning (RL) [12] are conducted to find out the best practice in eco-driving styles.

According to the Artificial Intelligence Index Report 2023 by Stanford University [13], academia led in releasing machine learning systems before 2014, and the industry has since dominated due to its greater capacity for data, computing power, and funding essential for advanced AI systems. Since such intelligent systems have high potential commercial value, enterprises usually consider protecting their intellectual property rights (IPR) by applying for patents in the target countries, e.g., the invention of data processing methods [14]. A patent is a legal grant of an exclusive right given to an inventor for a novel production process or a new method to address a technical problem. A patent grants an enterprise the exclusive right to prevent others from making, using, or selling similar inventions, and it serves to protect the interests of the patent assignee.

In general, patents indicate both the novelty and creativity of a product, as well as the commercialization potential of the inventor. Therefore, when conducting research in the engineering domain, specifically AI applications, it is essential to analyze not only scientific literature but also patents related to the research topic. Patents can provide valuable insights into existing technologies and techniques, allowing researchers to build on existing knowledge and develop new and innovative ideas. In addition, by reviewing patents, researchers can identify potential conflicts and modify their research accordingly, as well as identify gaps in the market that their research can fill. This analysis of patents can guide engineering researchers in upgrading their outputs into commercial operations, bridging the gap between academia and industry. Consequently, researchers can enhance the applicability and market viability of their work.

However, although there are literature reviews on eco-driving intelligent systems, no literature has covered patents in this domain. To fill the research gap, this paper conducts a patent search and review on eco-driving intelligent systems and AI algorithms. The patents classifications, priority dates and authorized patent offices are analyzed in this article, and different eco-driving intelligent systems and algorithms in the inventions are also discussed, as well as the differences of patents and articles.

The rest of the paper is organized as follows. The methodology and research process of patent review are presented in section II. Section III introduces the patent search results and the corresponding analysis. Section IV discusses the key aspects generated from the patents and the strengths and drawbacks of the study. Section V concludes the findings of the research and points out future research directions.

## II. METHODOLOGY OF PATENT SEARCH

This paper aims to investigate patents related to intelligent systems for eco-driving analysis, energy-consuming detection, and driving-style recommendation. In this paper, Google Patent database (https://patents.google.com/) is utilized to conduct a patent search. Google Patent is a comprehensive free database that includes both granted patents and published patent applications from hundreds of patent offices worldwide [15].

The patent search is performed in April 2023, and there are no limitations on the patent office and status in this paper. The search string (as shown in TABLE I. ) is designed and searched in the Google Patent Search system.

TABLE I. TABLE TYPE STYLES

| | |
|---|---|
| Search string | TI=(("Eco-driving" OR "Energy-efficient driving" OR "Green driving" OR "Fuel-efficient driving" OR "Sustainable driving" OR "Smart driving" OR "Driving behavior optimization" OR "Driving performance improvement" OR "Fuel consumption reduction" OR "Vehicle energy management")) |
| Note | "TI" represents that the keyword search is only conducted on titles due to many patents in the eco-driving field, which reduces the irrelevant results. |

Based on the search string on titles, 424 patents are found relevant [16]. There are three dates in a patent, including the priority date, filing date, and publication date. In this research, the priority date is a crucial element considered due to its role in determining the assignee's priority in claiming the invention, which can have a significant impact on the patent's novelty and inventive step. In the following sections, the related classifications, priority dates, and the authorized countries of the analyzed patents are discussed in detail.

## III. RESULTS

In this section, the patent search results are disclosed and analyzed. Section A introduces the patent classifications based on the Cooperative Patent Classification (CPC) system. The priority dates and the patent offices are discussed in section B. Finally, section C analyzes state-of-the-art patents filed after 2016 in detail.

### A. Cooperative Patent Classification

The CPC is a patent classification system developed by the European Patent Office (EPO) and the United States Patent and Trademark Office (USPTO) [17]. The CPC system is the specific and detailed extension of the International Patent Classification (IPC) system, and each CPC term is composed of a hierarchical structure including sections, classes, subclasses, and groups. For instance, the first letter in the symbol "B60W10/02" represents the section "Operations and Transport", which is followed by a two-digital number to denote the class ("B60" here refers to "vehicles in general"). The fourth character "W" represents the sub-classes involving conjoint control of vehicle sub-units of different types or different functions, control systems specially adapted for hybrid vehicles, or road vehicle drive control systems for purposes not related to the control of a particular sub-unit. The following group number "10/02" denotes conjoint control of vehicle sub-units of different types or different functions including control of driveline clutches.

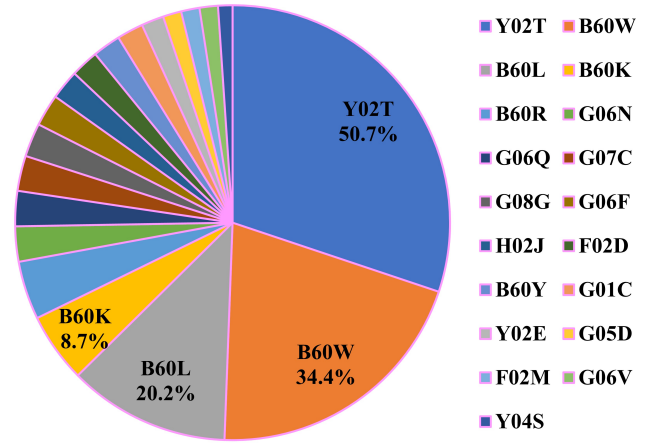

Fig. 1. Distribution of patent classifications based on CPCs.

Fig. 1 illustrates the distribution of the searched 424 patents' CPCs. The summation of their proportions is larger than 1 because each of them owns multiple CPC labels in the patent system. More than half of them is labeled as "Y02T", which denotes "climate change mitigation technologies related to transportation". The rest top-5 CPC classes are all "B60" which represents "vehicles in general". Such distribution shows that the majority of the inventions locate in the transportation sector and focus on ecological technologies.

TABLE II. shows the full names of the CPC classes of the related patents in this study, where the rankings are based on the proportions in the searched patents.

TABLE II. FULL NAMES OF RELATED CPCS

| CPCs | Full names |
|---|---|
| Y02T | Transportation-related climate change technologies |
| B60W | Conjoint control of diverse vehicle sub-units |
| B60L | Propulsion of electrically-propelled vehicles |
| B60K | Vehicle propulsion and transmission arrangement |
| B60R | Vehicles, vehicle fittings, or vehicle parts |
| G06N | Computing arrangements based on specific models |
| G06Q | ICT tailored for administrative, commercial, financial, managerial, or supervisory applications |
| G07C | Time or attendance registers |
| G08G | Traffic control systems |
| G06F | Electric digital data processing |
| H02J | Power supply or distribution circuit arrangements or systems |
| F02D | Controlling combustion engines |
| B60Y | Indexing scheme relating to aspects cross-cutting vehicle technology |
| G01C | Measuring distances, levels, or bearings |
| Y02E | GHG reduction in energy transmission, or distribution |
| G05D | Systems for controlling or regulating non-electric variables |
| F02M | Supplying combustible mixtures to combustion engines |
| G06V | Image or video recognition or understanding |
| Y04S | Integration of power network operation, communication, and information technologies to improve electrical power systems. |

## B. Priority Dates and Patent Offices

In most countries, the duration of patent rights for inventions is 20 years, so all patents before 2003 have expired. Moreover, deep learning has started to be popular since 2016, resulting in more deep learning-based intelligent systems in patent applications, and they deserve more in-depth research. As Fig. 2 demonstrates, there are 30 expired patents whose priority dates are before 2003 in the search results, and 219 latest patents filed after 2016. The sharp increase in patent applications over the past 20 years highlights the impact of advancements in AI and deep learning technologies on the development of eco-driving intelligent systems. Furthermore, the increasingly stringent environmental regulations and laws spur companies and research institutes to focus on developing energy-efficient driving strategies.

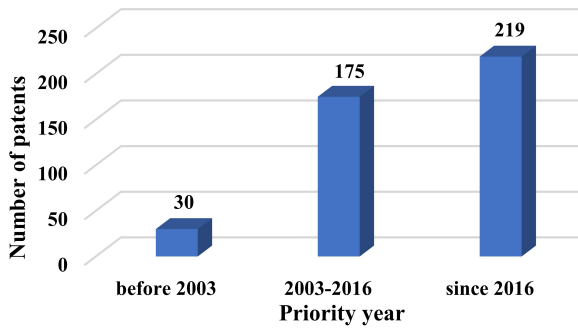

Fig. 2. Number of patents in distinct years.

Fig. 3 illustrates the patent distribution in different countries or regional patent offices where they are filed, granted, or published. 194 patents are processed by China National Intellectual Property Administration (CNIPA), and 157 of them are filed after 2016. Japan, South Korea, the United States, and Germany are other top patent offices dealing with eco-driving patents, and these 6 organizations make up approximately 88% of patents. The patents in Japan, South Korea, and Germany are concentrated between 2003 and 2016, whereas the US patent applications are stable since 2003.

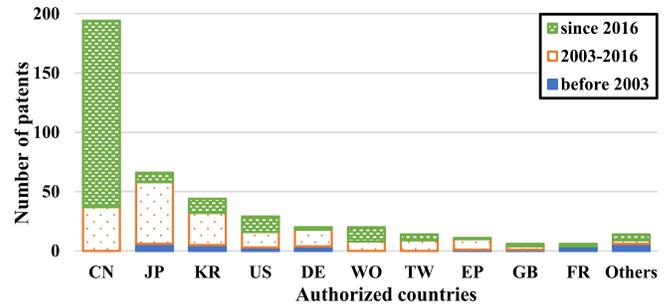

Fig. 3. Number of patents in different authorized patent offices. X-axis shows the country or re-gional patent office codes: CN (China), US (the USA), KR (South Korea), WO (World Intel-lectual Property Organization), JP (Japan), TW (Taiwan), DE (Germany), FR (France), GB (the UK), EP (European Patent Office).

Since deep learning is believed as the most promising state-of-the-art method, this paper mainly focuses on the 219 patents registered after 2016 and discusses them in detail in the following section.

## C. State-of-the-Art Patent Analysis

Fig. 4 demonstrates two different classification methods to categorize the 219 patents registered after 2016. Fig. 4(A) classifies the patents into five different groups based on the inventions' objectives and functions, whereas Fig. 4(B) categorizes them based on their technical nature.

The analyzed patents can be classified into the following five different groups:

- Vehicle energy management
- Smart driving
- Ecological and sustainable driving
- Fuel consumption reduction
- Driving behavior optimization

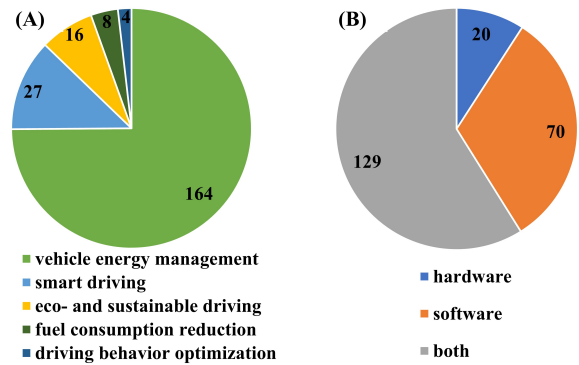

Fig. 4. Distribution of patents in different categories. (A) categories the patents based on their objectives and functions. (B) categories them based on their technical nature.

Fig. 5 illustrates the trend of patent applications of each category since 2016. Patents related to "vehicle energy management" dominate the applications, while the number of patents in the other four groups decreases after 2018. This trend indicates that energy management systems have become a popular research topic. The increasing popularity can be attributed to the rapid development of electric and hybrid vehicles, which is a response to stringent emission regulations implemented in recent years. Moreover, publishing a patent application requires up to 18 months, leading to a decrease in the availability of open documentation after 2021.

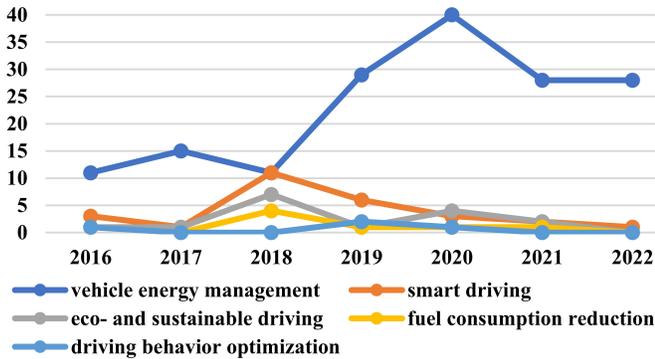

Fig. 5. Trend of each patent category since 2016.

As shown in Fig. 4(A), the distribution of patents among the above five categories. 74.9% of the inventions are related to vehicle energy management. This is likely because of the recent development of electric vehicles (EVs), where the battery's energy management is a critical research topic. An efficient energy management system optimizes the utilization of energy within a vehicle and reduce the environmental impact by reducing the GHG emissions. The vehicle energy management system in [18] includes a vehicle energy coordination unit to predict the energy consumption in the future route and optimize speed based on the traffic data. The predictive energy management system in [19] provides an energy-efficient driving scheme that takes into account the user's customized energy supplementing requirement.

Inventions in the smart driving category primarily disclose intelligent driving assistance systems and data-processing algorithms. Despite that some patents are not directly related to energy efficiency and ecological driving, some claims made in these publications can be extended to include such aspects. The smart driving control method in [20] adopts a framework to detect whether the current vehicle operating state matches the traffic signal. This method can be extended to incorporate eco-driving constraints, such as speed and fuel consumption limitations, making it suitable for eco-driving applications. An object detection algorithm and device are documented in [21], encompassing a sensor network for collecting camera and radar data, along with a neural network algorithm for detecting one or more targets. The objective of this system is to enhance driving energy efficiency through improved object detection capabilities.

Patents related to eco- and sustainable driving mainly concentrate on utilizing driving behavior information, vehicle status, and energy consumption predictions to enhance vehicle control strategies for improved energy efficiency. Many inventions [22, 23] disclose a comprehensive system for analyzing and decision-making, which includes data collection devices, data fusion and processing components, controllers, and decision-making algorithms. Such a system aims to optimize the energy efficiency of vehicles by providing effective and reliable decision-making support.

The reduction of fuel consumption and improvement of energy efficiency represents another important aspect found in the analyzed patents. Various devices [24] designed to detect fuel consumption and evaluate fuel efficiency are included in the patent documents. Additionally, some patents describe inventions that limit fuel consumption through some control strategies [25].

Finally, four of the analyzed patents are related to driving behavior optimizations. It is a popular and significant topic in eco-driving; however, this topic only represents less than 2% of the analyzed patents, possibly because most of the research in this field is published as articles. The research in this area primarily focuses on state-of-the-art ML/DL algorithms to analyze ecological driving styles and identify the best practices for energy consumption [3]. Deep reinforcement learning (DRL) algorithm is often implemented in the driving behavior decision making system as well [26, 27].

The inventions can also be classified based on their technical nature into three groups: hardware, software, and a combination of both. Fig. 4(B) demonstrates that 58.9% of the analyzed inventions involve both hardware systems and data processing methods.

The first category, "Hardware", refers to inventions that focus on physical components of the intelligent system, such as circuits, sensors, data processing components, and assembly systems. For example, a patent from CNIPA describes an electric vehicle energy management system [28] that consists of an EV with a battery, power adapters, monitors, and controllers, as well as a power management system with a battery, power adapter, and controllers. A French patent [29] documents a fuel consumption limitation device that utilizes a motorization and the accelerator pedal position setpoint to restrict fuel consumption.

Secondly, "Software" encompasses energy management methods relying on machine learning (ML) or deep learning (DL) techniques. An example outlined in [30] is a vehicle energy management method based on deep reinforcement learning, which pre-processes the vehicle information, the slope, and other working condition data, and then constructs a DRL environment based on the processed data. The eco-driving method in [31] adjusts the speed gradient of a target speed profile for a vehicle. This adjustment is made based on the driver's request for the vehicle's speed gradient and the speed profile of the embedded eco-driving mode.

Lastly, the category, "both", pertains to patents that include both a hardware system and a data processing method. For instance, both a method of optimizing the speed for energy efficiency and a vehicle system comprising multiple processors are introduced in [32]. Moreover, the vehicle energy management system in [33] encompasses both hardware components and control algorithms to improve the overall fuel efficiency and performance in the trucking industry.

IV. DISCUSSION

This patent review comprehensively analyzes all eco-driving patents searched from the Google Patents database and highlights 219 cutting-edge inventions filed after 2016. The study reveals a rapid increase in patent applications within this field, driven by the advancement of AI technologies and more stringent environmental regulations. Fig. 3 demonstrates that China has dominated patent applications since 2016, possibly due to the significant growth of the Chinese vehicle market. As a result, both domestic and international corporations and research institutions should take steps to protect their intellectual property rights in China.

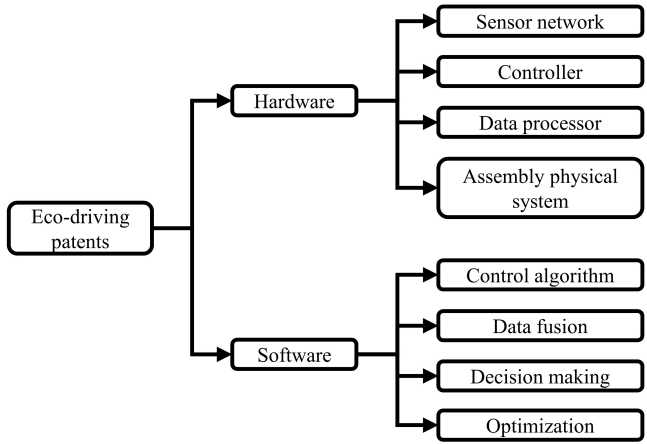

Fig. 6. Technology fields of inventions.

The 219 latest patents can be classified into five groups based on their topics as outlined in section III.C. All these categories encompass both hardware and software inventions. Fig. 6 illustrates the technology fields associated with the inventions concerning eco-driving intelligent systems. All inventions in filed patents consist of hardware and/or software.

The hardware inventions include data-collection sensors that detect environmental information, driving behavior information, and vehicle operating data. Certain inventions also publish controllers which operate specific control algorithms to enable the vehicles to operate in energy-efficient mode. Furthermore, the data processors in the filed patents are employed to execute the developed algorithms, which are responsible for processing the data gathered from the sensor networks or the vehicle embedded system such as CANBus or OBD. Lastly, many patents claim the entire intelligent system encompassing all mentioned physical components.

The software part in the inventions involves both novel control strategies and state-of-the-art AI algorithms. The control methods leverage mathematical and dynamic principles to regulate fuel consumption during vehicle operation, whereas the remaining three categories are all data-driven and AI-based algorithms. Data fusion algorithms are predominantly utilized to integrate information from collected data, removing redundant information and enhancing the reliability of data processing. Additionally, various ML/DL algorithms are developed to determine optimal driving styles, control parameters, and predict energy consumption for informed decision-making. Lastly, certain patents feature simulation frameworks based on reinforcement learning, wherein driving behaviors and vehicle operating styles are optimized in different simulation scenarios.

Specifically, the hardware components of the inventions falling under the 'both' category in Fig. 4(B) primarily pertain to the physical assembly system. This system involves devices that facilitate data collection, processing, control, and visualization. Such inventions also incorporate various data processing algorithms, including data preprocessing techniques, prediction models, optimization algorithms, and decision-making frameworks. For instance, the eco-driving assistance system in [22] comprises two information collection devices, an energy consumption prediction device, and a vehicle operating decision device. On the software front, this invention introduces a driving style classification algorithm, a data fusion algorithm that combines operating information and best practices, and a decision-making framework to enhance eco-driving performance.

In general, the analyzed patents in this paper mainly focus on the entire intelligent systems including both the devices and data processing algorithms, as Fig. 4(B) shows that 58.9% of the inventions involve both novel hardware and software. In addition, only a few deep learning algorithms are mentioned in the patent files. This could be because the majority of novel algorithms are typically published in the form of scientific articles, and patent laws generally request that models published in other forms cannot be patented.

In summary, this study provides general and preliminary results by analyzing patent titles, CPCs, and countries. To obtain more specific findings, an in-depth analysis of patent claims is needed to define the boundaries of the inventions. Such analysis will facilitate comprehensive recommendations for patent applications in the domain of eco-driving intelligent systems. Moreover, patents should be compared with scientific articles, exploring the difference between academic frameworks and industry systems.

V. CONCLUSIONS

This article provides an overview of 424 patents filed for eco-driving intelligent systems from the Google Patent database. The discussion covers general patent information, such as CPCs, priority dates, and patent offices, and different categories of patents filed after 2016 are analyzed in detail. The results show that 50.7% of the patents are labeled as "Y02T", which refers to "climate change mitigation technologies related to transportation", and the remaining top-5 CPC classes are all "B60 - vehicles in general". Additionally, the number of patent applications has increased rapidly since 2016, with over 70% of them granted in CNIPA. Furthermore, 219 patents filed after 2016 are categorized into vehicle energy management, smart driving, eco- and sustainable driving, fuel consumption reduction, and driving behavior optimization. These categories encompass physical components of intelligent systems and novel ML/DL algorithms.

However, this patent review provides only a general and preliminary analysis of patents. Therefore, an in-depth analysis of the patents, especially their claims, is recommended for future work, to make comprehensive recommendations for commercial-level production development related to eco-driving intelligent systems. In conclusion, eco-driving intelligent systems should be implemented to reduce energy consumption in transportation systems, and the corresponding intellectual property rights should be protected through patent applications. The commercialization potential should also be investigated before documenting a patent. It is crucial to consider these findings and recommendations while making decisions about patent applications in this field.

ACKNOWLEDGMENT

This paper is part of the project "Driver Coach" (Case no. 64021-2034) by the Danish funding agency, the Danish Energy Technology Development and Demonstration (EUDP) program, Denmark.